\documentclass[12pt]{article}
\RequirePackage[authoryear]{natbib}

\RequirePackage{amsthm,amsmath,amsfonts,amssymb}
\RequirePackage[colorlinks,citecolor=blue,urlcolor=blue]{hyperref}
\RequirePackage{graphicx}

\usepackage{bm}
\usepackage{amsmath}
\usepackage{amssymb}
\usepackage{natbib}
\usepackage{graphics}
\usepackage{enumitem}

\usepackage[plain,noend]{algorithm2e}
\usepackage{color}
\usepackage{multirow}
\usepackage{float}

\usepackage{url}
\usepackage{bbm}
\usepackage[margin=1in]{geometry}
\usepackage{subcaption}
\usepackage{caption}

\def\ESE{\textsc{ese}}
\def\RMSE{\textsc{rmse}}
\def\BIAS{\textsc{bias}}
\def\CP{\textsc{cp}}
\def\LEN{\textsc{len}}
\def\BMI{\textsc{bmi}}
\def\CHD{\textsc{chd}}
\def\CHF{\textsc{chf}}

\newcommand{\bA}{\boldsymbol{A}}
\newcommand{\ba}{\boldsymbol{a}}

\newcommand{\bB}{\boldsymbol{B}}

\newcommand{\bC}{\boldsymbol{C}}

\newcommand{\bD}{\boldsymbol{D}}
\newcommand{\mD}{\mathcal{D}}

\newcommand{\bH}{\boldsymbol{H}}

\newcommand{\bI}{\boldsymbol{I}}

\newcommand{\bM}{\boldsymbol{M}}

\newcommand{\bR}{\boldsymbol{R}}

\newcommand{\bS}{\boldsymbol{S}}

\newcommand{\bU}{\boldsymbol{U}}

\newcommand{\bV}{\boldsymbol{V}}

\newcommand{\bW}{\boldsymbol{W}}

\newcommand{\bX}{\boldsymbol{X}}
\newcommand{\bx}{\boldsymbol{x}}

\newcommand{\by}{\boldsymbol{y}}

\newcommand{\bzero}{\boldsymbol{0}}
\newcommand{\balpha}{\boldsymbol{\alpha}}
\newcommand{\bbeta}{\boldsymbol{\beta}}
\newcommand{\tbeta}{\widetilde{\boldsymbol{\beta}}}

\newcommand{\bmu}{\boldsymbol{\mu}}

\newcommand{\bpsi}{\boldsymbol{\psi}}
\newcommand{\bSigma}{\boldsymbol{\Sigma}}

\allowdisplaybreaks

\newtheorem{theorem}{Theorem}

\theoremstyle{definition}

\allowdisplaybreaks

\begin{document}

\title{\bf Statistical Inference for Streamed Longitudinal Data}
\author{Lan Luo\\
Department of Statistics and Actuarial Science, University of Iowa\\
Jingshen Wang\\
Division of Biostatistics, University of California, Berkeley\\
Emily C. Hector\\
Department of Statistics, North Carolina State University}
  \date{}
\maketitle
 
\bigskip

\begin{abstract}
Modern longitudinal data, for example from wearable devices, measures biological signals on a fixed set of participants at a diverging number of time points. Traditional statistical methods are not equipped to handle the computational burden of repeatedly analyzing the cumulatively growing dataset each time new data is collected. We propose a new estimation and inference framework for dynamic updating of point estimates and their standard errors across serially collected dependent datasets. The key technique is a decomposition of the extended score function of the quadratic inference function constructed over the cumulative longitudinal data into a sum of summary statistics over data batches. We show how this sum can be recursively updated without the need to access the whole dataset, resulting in a computationally efficient streaming procedure with minimal loss of statistical efficiency. We prove consistency and asymptotic normality of our streaming estimator as the number of data batches diverges, even as the number of independent participants remains fixed. Simulations highlight the advantages of our approach over traditional statistical methods that assume independence between data batches. Finally, we investigate the relationship between physical activity and several diseases through the analysis of accelerometry data from the National Health and Nutrition Examination Survey.
\end{abstract}

\noindent%
\vfill

\section{Introduction}
\label{s:introduction}

Traditionally, longitudinal studies have collected a moderate number of observations from a large number of individuals, with the ultimate goal of drawing statistical inference for this ever-growing set of participants. With the advent of modern technologies such as smartphones and wearable devices, the data collection paradigm has shifted to an infinite horizon setting in which data is collected on a fixed number of participants in perpetuity. This new setting offers the possibility of discovering new patterns in human daily life specific to a set of individuals, opening the door to targeted biomedical interventions for population subgroups.

A typical example consists of wearable device data that is frequently uploaded to a smartphone application and summarized into a few health metrics. A biomedical research question might focus on the relationship between physical activity and covariates such as phenotype across a fixed set of users, and pose a parametric model to study this relationship. Each time users upload new data to the application, new data becomes available to answer this research question, but the memory and computational burdens of storing and analyzing the entire cumulative dataset, consisting of all observations prior to the latest upload, is astronomical. Instead of reanalyzing the entire dataset each time new data is available, it is preferable to update parameter estimates from previous batches of data with the new batch in a computationally efficient approach. 

Numerous statistical and computational challenges arise when considering intensively measured longitudinal data on a limited set of participants. The sheer size and complexity of the data frequently prohibit statistical analyses of the entire data due to computational or modelling challenges. Distributed and online approaches have gained popularity in the statistics literature as viable alternatives to whole-data approaches, with numerous solutions proposed for independent and dependent data alike \citep{XieCD2013, Zhou-Song-2017, Jordan2019CSL,  Hector-Song-JASA, Duan2021}. Technically speaking, the challenge in infinite horizon settings is to derive updating rules for point estimates as well as measures of uncertainty, such as standard errors, across dependent batches of data. The online paradigm is a natural framework for processing data batches that are collected serially; it also presents a technical advantage over the distributed computing paradigm: while both only require the storing of summary statistics, the distributed setting requires the number of participants in each data batch to be large, an assumption that is clearly violated in our infinite horizon setting with a finite number of participants.

The majority of research efforts in developing procedures that allow for quick updates of parameter estimates fall into the field of online learning. This line of research dates back five decades to when \cite{Robbins1951} proposed a stochastic approximation algorithm that laid a foundation for the popular stochastic gradient descent algorithm \citep{Sakrison1965}. The stochastic gradient descent algorithm and its variants have been extensively studied for online estimation and prediction \citep{Toulis2017}, but the work of developing online statistical inference remains unexplored. Recently, \cite{Fang2019} proposed a perturbation-based resampling method to construct confidence intervals for stochastic gradient descent, but it does not achieve desirable statistical efficiency and may produce misleading inference in the case of large regression parameters. In addition to the stochastic gradient descent types of recursive algorithms, several cumulative updating methods have been proposed to specifically perform sequential updating of regression coefficient estimators, including the online least squares estimator for the linear model \citep{optimal1994, Chen2006linear}, the cumulative estimating equation estimator and the cumulatively updated estimating equation estimator by \cite{Schifano2016CUEE} for nonlinear models. It is worth noting that both the cumulative estimating equation and the cumulatively updated estimating equation are developed under a mechanism similar to meta-analysis, and estimation consistency is established under a strong regularity condition that the number of data batches is much smaller than the sample size of each data batch \citep{lin2011aggregated, Schifano2016CUEE}. Recently, \cite{Luo-Song-2020} proposed a renewable estimation and incremental inference method that is asymptotically equivalent to the maximum likelihood estimators obtained from the entire dataset. More importantly, this method overcomes the unnatural constraint on the number of data batches versus the sample size of each data batch. It does not, however, allow for dependence between the data batches.

Indeed, most of the aforementioned online algorithms are developed under the assumption that samples collected at different time points are independently generated from the same underlying model. A prominent concern in mobile health data analyses is that there exists a non-negligible degree of correlation across different sampling points. Ignoring this correlation may not affect estimation consistency, but it may lead to a loss of statistical efficiency and therefore produce misleading inference in real-time decision-making. In the presence of non-trivial dependence between observations in streaming data, \cite{Cappe2011} proposed an online expectation-maximization algorithm for parameter estimation in hidden Markov models, but did not provide a way for inference. To the best of our knowledge, real-time inference with dependent data batches remains largely unexplored. Unlike the independent setting, real-time statistical inference cannot be done through a simple linear aggregation of inferential quantities such as information matrices like in \cite{Luo-Song-2020}. Instead, the within-individual correlation induces a huge non-diagonal covariance matrix of dimension $n\times n$, where $n$ is the number of cumulative repeated measurements per individual that grows rapidly over time. In addition to the nontrivial challenge of incorporating dependence between data batches, existing methods impose a parametric model with a fixed, common parameter that does not change over time. This assumption is too restrictive for high-throughput data collection. For example, the association between explanatory variables such as sex and outcomes such as physical activity collected from wearable devices is expected to change dynamically over time rather than remain constant. Failing to account for these intrinsic dynamics can lead to severely misleading estimation and inference. A model that can handle local time dynamics is highly desirable, but lacking in the current literature \citep{Schifano2016CUEE,Luo-Song-2020}

In this paper, we propose a new framework for online updating of point estimates and uncertainty quantification for time-varying associations in infinite horizon longitudinal data settings. We derive a new result that shows how the extended score function of the quadratic inference function of \cite{Qu-Lindsay-Li} constructed over the cumulative longitudinal data elegantly decomposes into a sum of summary statistics over data batches. We further demonstrate how it can be computed recursively using only the summary statistic from the previous data batch, a formulation which lends itself naturally to online updating over dependent data batches. To account for local dynamics, we incorporate an exponential weight function to dynamically adjust the weights applied to historical data batches. This approach is unique in that, in contrast to existing nonparametric approaches, it uses a one-sided kernel that only uses data prior to the current observations for weighting. We solve the substantial technical challenge of deriving the asymptotic convergence rate of our online dynamic estimator under the infinite horizon longitudinal data setting with finite sample size. In doing so, we move away from traditional assumptions of independence, and embrace new frameworks that reflect local dynamics in streaming data environments. As a special case, our proposed approach can be adopted for individual-level analysis. The resulting proposed method leverages the dependence between data collected sequentially on the same set of participants to estimate dynamic associations with improved statistical efficiency.

\section{Streaming Inference for Longitudinal Data}
\label{sec:method}

\subsection{Problem setup}
\label{ss:setup}

For scalars $s_1\ldots, s_{\ell}$, vectors $\ba_{\ell} \in \mathbb{R}^{v}$ and matrices $\bA_{\ell} \in \mathbb{R}^{q_{\ell}\times v}$, $\ell=1, \ldots, L$, define the column stacking operation on scalars, vectors and matrices as 
\begin{equation*}
\begin{split}
(s_{\ell})_{\ell=1}^L = \begin{pmatrix}
s_1 & \ldots & s_L
\end{pmatrix}^\top \in \mathbb{R}^L, \quad
( \ba^\top_{\ell} )_{\ell=1}^L = \begin{pmatrix}
\ba_1 & \ldots & \ba_L
\end{pmatrix}^\top \in \mathbb{R}^{L \times v},\\
( \bA_{\ell} )_{\ell=1}^L = \begin{pmatrix}
\bA^\top_1 & \ldots & \bA^\top_L
\end{pmatrix}^\top \in \mathbb{R}^{\sum_{\ell=1}^L q_{\ell} \times v},~~~~~~~~~~~~~~~~~~~~
\end{split}
\end{equation*}
respectively. Suppose we collect data batches $\mD_{i,j}= \{ \by_{i,j}, \bX_{i,j} \}$ sequentially at deterministic updating time points $t_j,\ j=1,2,\dots,b$, on the same set of independent participants $i=1,\ldots, m$, where $\by_{i,j} \in \mathbb{R}^{n_j}$ is the multivariate longitudinal outcome in batch $j$ and $\bX_{i,j} = ( \bx^\top_{i,kj} )_{ k=1 }^{ n_j } \in \mathbb{R}^{n_j \times p}$ is a matrix of $p$ explanatory variables and $n_j$ repeated measurements with $\bx_{i,kj}=(x_{i,1kj}, \ldots, x_{i,pkj}) \in \mathbb{R}^p$, $k=1, \ldots, n_j$, $j=1, \ldots, b$. Let $\mD_{i,b}^\star=\{\mD_{i,1},\dots,\mD_{i,b}\}$ denote the cumulative dataset up to batch $b$ in participant $i$, and $N_b=\sum_{j=1}^b n_j$ denote the corresponding aggregated response dimension. For ease of exposition, we assume an equal batch size $n_j$ and equal number of repeated measurements $N_b$ for all participants. We consider the marginal generalized linear model with outcome $\by_i=(\by^\top_{i,1},\ldots,\by^\top_{i,b})^\top \in \mathbb{R}^{N_b}$ and $p$ covariates $\bX_i=(\bX_{i,j})_{j=1}^b \in \mathbb{R}^{N_b \times p}$:
\begin{align}
E(\by_i \mid \bX_i )=\bmu_i=(\mu_{i,kj})_{k,j=1}^{n_j,b} = \left\{ h\left( \bx_{i,kj}^\top \bbeta(t_j) \right) \right\}_{k,j=1}^{n_j,b} \in \mathbb{R}^{N_b}, \quad i=1,\dots,m,
\label{e:MGLM}
\end{align}
where $\bbeta(\cdot)$ is the regression coefficient function, and $h(\cdot)$ is a known link function. The regression coefficient $\beta(\cdot)$ is assumed to be a smooth function that captures local dynamics of the associations between outcomes and explanatory variables. We consider a batch-varying coefficient, denoted by $\beta(t_j)\in\mathbb{R}^p$, for the batch of data collected at time $t_j$. For notation simplicity, we use $\beta_j\in\mathbb{R}^p$ to denote the true value of the batch-specific coefficient, and $\beta$ as a generic notation for the coefficient function. 

The covariance of the outcome is given by $\text{cov} (\by_i \mid \bX_i) \propto \bA^{1/2}_i \bR( \balpha) \bA^{ 1/2 }_i$, where $\bA_i = \text{diag} \{ v(\mu_{i,kj})\}_{k,j=1}^{n_j,b}$ is a diagonal matrix with $v(\cdot)$ a known variance function, and $\bR(\balpha)$ is a working correlation matrix that is fully characterized by a correlation parameter $\balpha$. Clearly, the dimensions of $\by_i$, $\bX_i$, $\bA_i$ and $\bR(\cdot)$ depend on the number of batches $b$, which is allowed to diverge. We use the subscript $i$ to refer to the cumulative data on participant $i$ up to batch $b$; the dependence on $b$ is suppressed for parsimony of notation, but we give a reminder of this fact when relevant. To study the relationship between the longitudinal outcomes and covariates, we focus on estimation and inference for {$\bbeta$} with $\balpha$ treated as a nuisance parameter. 

Due to the longitudinal nature of the infinite horizon setting, we model the longitudinal measurements through a first-order autoregressive process. The first-order autoregressive process is one of the most widely used correlation models for longitudinal and time series correlations because it provides a natural description of the exponential decay in correlation between measurements that occurs over time. Specifically, this process assumes the correlation between two longitudinal outcomes $Y_{i,t}, Y_{i,s}$, measured at time points $t$ and $s$ respectively, satisfies a serial structure given by $corr(Y_{i,t}, Y_{i,s}) = \alpha^{\mid t-s\mid}$, $t\neq s$, $\alpha \in (-1,1)$. 

\subsection{Offline approaches for longitudinal data analysis}
\label{ss:offline}

Full likelihood-based estimation and inference for {$\bbeta$} depends on the specification of an $N_b$-dimensional multivariate likelihood and parametrization of all moments up to order $N_b$, which poses modelling and computational difficulties. While these prove trivial to overcome for multivariate Gaussian likelihoods, the specification of the likelihood for non-Gaussian outcomes typically relies on copulas \citep{Song, Joe-2}, which are computationally burdensome. Alternative quasi-likelihood based approaches, such as the composite likelihood \citep{Lindsay, Varin-Reid-Firth}, prove a necessary alternative, but frequently at the cost of statistical efficiency. The generalized estimating equations \citep{Liang-Zeger} avoid the need for a likelihood by directly specifying an estimating equation for $\bbeta$, with $\balpha$ estimated using a method of moments approach. Up to time $t_b$, a generalized estimating equations estimator of $\bbeta$ is the solution to the weighted generalized estimating equation based on data $\{\mD_{i,b}^\star\}_{i=1}^m$, denoted by $\bpsi_b^\star
(\bbeta, \balpha; \{\mD_{i,b}^\star\}_{i=1}^m ) = 
\sum_{i=1}^m \bD_i^\top\bSigma_i^{-1} \bW_b (\by_i - \bmu_i) = 
\bzero$,
where 
${\bD}_i = \nabla_{\bbeta} \bmu_i = \left(\bD_{i,j}\right)_{j=1}^b\in\mathbb{R}^{N_b\times p}$, and $\bSigma_i=\bA_i^{1/2}\bR(\balpha)\bA_i^{1/2}$ with ${\bA}_i =\text{diag} \{\bA_{i, j} \}_{j=1}^b\in\mathbb{R}^{N_b\times N_b}$. Here, we introduce an additional weighting matrix $\bW_b=\text{diag}\{\bW_{b,j}\}_{j=1}^b\in\mathbb{R}^{N_b\times N_b}$ to the original generalized estimating equation framework. This matrix dynamically adjust the weights assigned to data batches collected at different time points. In particular, we define
$\bW_{b,j} = q^{t_b-t_j} \bI_{n_j}, \ 0< q < 1$: with this weight function, observations in batches that are further away from batch $b$ receive less weights. We remind the reader that the dimensions of $\bD_i$, $\bA_i$, $\bW_b$, $\bR(\cdot)$ and $\bSigma_i$ depend on $b$, which is allowed to diverge. The generalized estimating equations has enjoyed widespread popularity in the analysis of longitudinal data for its ease of implementation and its desirable statistical properties: when the correlation structure is correctly specified by $\bR(\balpha)$, the generalized estimating equations estimator is consistent and semi-parametrically efficient, i.e. as efficient as the quasi-likelihood. Even when the correlation structure is misspecified, the generalized estimating equations estimator remains consistent. Unfortunately, there exist simple cases for which the estimator of the correlation parameter $\balpha$ does not exist \citep{Crowder}. Generally, estimation of $\balpha$ is cumbersome since the target of inference is $\bbeta$, and a preferable approach would bypass estimation of the correlation structure altogether. 

The quadratic inference function of \cite{Qu-Lindsay-Li} avoids estimation of $\balpha$ through a clever substitution of a linear expansion of known basis matrices for the inverse of the working correlation matrix in the generalized estimating equations. The formulation of the quadratic inference function is based on an approximation to the inverse of the working correlation matrix by $\bR^{-1}(\balpha)\approx \sum_{s=1}^S \gamma_s \bM_s$, where $\gamma_1,\dots,\gamma_S$ are unknown constants possibly dependent on $\balpha$, and $\bM_1,\dots,\bM_S \in \mathbb{R}^{N_b \times N_b}$ are known basis matrices with elements $0$ and $1$, which are determined by a given correlation structure in $\bR(\balpha)$. Plugging this expansion into the generalized estimating equations leads to $\bpsi_b^\star(\bbeta, \balpha; \{\mD_{i,b}^\star\}_{i=1}^m) = \sum_{i=1}^m\sum_{s=1}^S \gamma_s\bD_i^\top\bA_i^{-1/2}\bM_s\bA_i^{-1/2}\bW_b(\by_i - \bmu_i) = \bzero$, which can be expressed as a linear combination of the following extended score vector:
\begin{align}
\bU_b^\star(\bbeta) = \sum \limits_{i=1}^m \bU_{i,b}^\star(\bbeta) = \sum_{i=1}^m 
\begin{pmatrix}
\bD_i^\top\bA_i^{-1/2} \bM_1\bA_i^{-1/2}\bW_b(\by_i - \bmu_i) \\
\vdots \\
\bD_i^\top\bA_i^{-1/2} \bM_S\bA_i^{-1/2}\bW_b(\by_i - \bmu_i)
\end{pmatrix} \in \mathbb{R}^{pS}.
\label{e:QIF-offline-U}
\end{align}
Again, we introduce an additional weighting matrix $\bW_b$ to the original quadratic inference function framework to dynamically adjust the weights assigned to data batches collected at different time points.

Clearly, estimation of the second-order moments of $\by_i$ is no longer required in the quadratic inference function since $\gamma_1, \ldots, \gamma_S$ are not involved in the construction of $\bU_b^\star(\bbeta)$. The extended score function in \eqref{e:QIF-offline-U} is an over-identified estimating function: $pS= dim(\bU_b^\star(\bbeta)) \allowbreak > dim(\bbeta)=p$. To obtain an estimator of $\bbeta$, following \cite{Hansen}'s generalized method of moments, the quadratic inference function estimator is defined as $\widehat{\bbeta}_b^\star = \arg \min_{\bbeta} Q_b^\star(\bbeta)$ with
\begin{equation}\label{eq:QIF-offline-Q}
Q_b^\star(\bbeta) = {\bU_b^\star(\bbeta)}^\top \left\{{\bV}_b^\star(\bbeta)\right\}^{-1} {\bU}_b^\star(\bbeta),
\end{equation}
where $\bV_b^\star(\bbeta)=\sum_{i=1}^m \bU_{i,b}^\star(\bbeta)\bU_{i,b}^\star(\bbeta)^\top$ is the sample covariance matrix of $\bU_b^\star(\bbeta)$. The correlation parameter $\balpha$ is not involved in \eqref{eq:QIF-offline-Q}, so that the estimation of $\bbeta$ using $\widehat{\bbeta}_b^\star$ yields substantial computational gains over the generalized estimating equations when $N_b$ is large. A suitable approximation to the inverse of a first-order autoregressive working correlation structure is given by $\bR^{-1}(\balpha) \approx \gamma_1 \bM_1 + \gamma_2 \bM_2$ with $\bM_1=\bI_{N_b}$ the $N_b\times N_b$ identity matrix and $\bM_2$ a matrix with $1$ on the two main off-diagonals and $0$ elsewhere \cite{Qu2000}. While a third basis matrix is sometimes used to capture edge effects, using these two basis matrices for the first-order autoregressive process gives a satisfactory efficiency gain in practice \cite[Chapter 5]{Song2007correlated}. 

It is well known that the quadratic inference function estimator $\widehat{\bbeta}_b^\star$ is consistent even if the correlation structure imposed by the choice of $\bM_1, \ldots, \bM_S$ is misspecified, and that it is semi-parametrically efficient when the correlation structure is correctly specified \citep{Qu-Lindsay-Li}. In addition, it has been shown both theoretically and numerically that estimation efficiency of the quadratic inference function estimator $\widehat{\bbeta}_b^\star$ is higher than of the generalized estimating equations estimator under correlation misspecification \citep{Qu-Lindsay-Li, Song-etal}. Following standard generalized method of moments theory \citep{Hansen}, misspecification of the correlation structure does not impact estimation consistency but only estimation efficiency. The use of the quadratic inference function reduces the difficulty of computation on the cumulative data $\{ \mD^\star_{i,b}\}_{i=1}^m$ over likelihood and quasi-likelihood based approaches by avoiding estimation of the nuisance parameter related to second-order moments of the outcome. Nonetheless, it does not provide a satisfactory solution to the tremendous memory and computational costs incurred by the analysis of the cumulative data.

\subsection{A new decomposition for quadratic inference functions with data batches}

We derive a new result that shows how the extended score function of the quadratic inference function decomposes into a sum of summary statistics over batches of data. First, we partition the extended score vector $\bU_b^\star(\bbeta)$ into two subvectors based on the basis matrices: let ${\bU}^\star_b(\bbeta)^{(1)} \in\mathbb{R}^p$ denote the subvector corresponding to the identity basis matrix $\bM_1$ while ${\bU}^\star_b(\bbeta)^{(2)}\in\mathbb{R}^p$ denotes the subvector that involves $\bM_2$. The extended score function up to data batch $b$ can be written as
\[
{\bU}^\star_b(\bbeta) = \begin{pmatrix} 
{\bU}^\star_{b}(\bbeta)^{(1)}  \\
{\bU}^\star_{b}(\bbeta)^{(2)}
\end{pmatrix}
=\sum_{i=1}^m
\begin{pmatrix}
{\bU}_{i,b}^\star(\bbeta)^{(1)} \\
{\bU}_{i,b}^\star(\bbeta)^{(2)}
\end{pmatrix}
=\sum_{i=1}^m
\begin{pmatrix}
{\bD}_i^\top {\bA}_i^{-1/2} \bM_1 {\bA}_i^{-1/2} \bW_b ({\by}_i - {\bmu}_i) \\
{\bD}_i^\top {\bA}_i^{-1/2} \bM_2 {\bA}_i^{-1/2} \bW_b ({\by}_i - \bmu_i)
\end{pmatrix}.
\]

Since $\bM_1$ is an identity matrix of dimension $N_b\times N_b$ that is equivalent to the case of independent data batches, $\bU_b^\star(\bbeta)^{(1)}=\sum_{i=1}^m\sum_{j=1}^bq^{t_b-t_j}\bU_{i,j}(\bbeta)^{(1)}$ is a linear aggregation of the score functions corresponding to each data batch $\mD_{i,j},\ j=1, \dots, b$ for each subject $i=1,\dots,m$. Decomposition of $\bM_2$ over data batches, however, is not trivial. 
Let $\bM_{2,j}\in\mathbb{R}^{n_j\times n_j}$ denote the diagonal blocks of $\bM_2$ corresponding to the $j$th data batch, and $\bB_{j}\in\mathbb{R}^{n_j\times n_{j+1}}$ be the off-diagonal blocks with $1$ in the $(n_j,1)$-entry and $0$ elsewhere. Then $\bU_b^\star(\bbeta)^{(2)}$ decomposes as
\begin{align}\label{eq:offlineEE-pieces}
&\sum_{i=1}^m \bU_{i,b}^\star(\bbeta)^{(2)}
= \sum_{i=1}^m \bD_i^\top \bA_i^{-1/2} \bM_2 \bA_i^{-1/2} W_b(\by_i - \bmu_i) \nonumber \\
& = \sum_{i=1}^m 
\begin{pmatrix}
\bD_{i,1} \\
\vdots \\
\bD_{i,b}
\end{pmatrix}^\top
	\begin{pmatrix}
	\bA_{i,1}& &\\
	&\ddots & \\
	& & \bA_{i,b}
	\end{pmatrix}^{-1/2}
	\begin{pmatrix}
	\bM_{2,1}&\bB_{1} &\\
	&\ddots &\\
	&\bB_{b}^\top & \bM_{2,b}
	\end{pmatrix}
    \begin{pmatrix}
    \bA_{i,1}& &\\
    &\ddots & \\
    & & \bA_{i,b}
    \end{pmatrix}^{-1/2}\\
&~~~~~~~~~~~~~~~~~~~~~~~~~~~~~~~~~~~~~~~~~~~~~~~~~~~~~~~~~~~~~~~~~~~~~~~~~~~~~~~~~~~~~~~~~~~~~~~
    \begin{pmatrix}
	\bW_{b,1}(\by_{i,1}-\bmu_{i,1}) \\
	\vdots \\
    \bW_{b,b}(\by_{i,b}-\bmu_{i,b})
	\end{pmatrix} \nonumber \\
&=\sum_{i=1}^m  \sum_{j=1}^{b} q^{t_b-t_j} \bU_{i,j}(\bbeta)^{(2)} +\sum_{i=1}^m \sum_{j=1}^{b-1} q^{t_b-t_{j+1}} \bU_{i,j,j+1}(\bbeta) +\sum_{i=1}^m \sum_{j=1}^{b-1} q^{t_b-t_j} \bU_{i,j+1,j}(\bbeta),
\end{align}
with 
\[
\begin{split}
\bU_{i,j}(\bbeta)^{(1)} &= \bD_{i,j}^\top \bA_{i,j}^{-1/2} \bM_1 \bA_{i,j}^{-1/2}(\by_{i,j}-\bmu_{i,j}), \\
\bU_{i,j}(\bbeta)^{(2)} &=\bD_{i,j}^\top \bA_{i,j}^{-1/2} \bM_2 \bA_{i,j}^{-1/2}(\by_{i,j}-\bmu_{i,j}), \\
\bU_{i,j,j+1}(\bbeta) &= \bD_{i,j}^\top \bA_{i,j}^{-1/2} \bB_{j}\bA_{i,j+1}^{-1/2}(\by_{i, j+1}-\bmu_{i, j+1}), \\
\bU_{i,j+1,j}(\bbeta)&=\bD_{i,j+1}^\top\bA_{i,j+1}^{-1/2} \bB_j^\top
\bA_{i,j}^{-1/2}(\by_{i,j}-\bmu_{i,j}).
\end{split}
\]

The corresponding negative gradient matrices are denoted by 
\[
\begin{split}
    \bS_{i,j}(\beta)^{(1)} &= \bD_{i,j}^\top \bA_{i,j}^{-1/2} \bM_{1,j}\bA_{i,j}^{-1/2} \bD_{i,j}, \quad~~~ \bS_{i,j}(\beta)^{(2)} =\bD_{i,j}^\top \bA_{i,j}^{-1/2} \bM_{2,j} \bA_{i,j}^{-1/2} \bD_{i,j},\\
    \bS_{i,j,j+1}(\beta) &=\bD_{i,j}^\top \bA_{i,j}^{-1/2} \bB_j \bA_{i,j+1}^{-1/2} \bD_{i,j+1}, \quad \bS_{i,j+1,j}(\beta) =\bD_{i,j+1}^\top \bA_{i,j+1}^{-1/2} \bB_j^\top \bA_{i,j}^{-1/2} \bD_{i,j}.
\end{split}
\]

Thus, we have shown that $\bU^\star_b(\bbeta)^{(2)}$ elegantly decomposes into estimating functions for batch-specific (through $\bU_{i,j}(\bbeta)^{(2)}$) and inter-batch dependencies (through $\bU_{i,j,j+1}(\bbeta)$ and $\bU_{i,j+1,j}(\bbeta)$). This decomposition effectively breaks down the massive correlation matrices of dimension $N_b\times N_b$ into smaller matrices of dimension $n_j\times n_j$. While this construction seems more appealing, we have not yet reduced the computational and memory difficulties associated with $Q_b^\star(\bbeta)$ in \eqref{eq:QIF-offline-Q}. Indeed, \eqref{eq:QIF-offline-Q} must be solved each time a new data batch arrives and it depends on the cumulative data $\{\mD_{i,b}^\star \}_{i=1}^m$ up to batch $b$. In Section \ref{ss:online}, we show how to use the decomposition in \eqref{eq:offlineEE-pieces} to avoid processing the cumulative data.

\subsection{The streaming inference framework}
\label{ss:online}

Instead of processing the cumulative dataset $\{\mD_{i,b}^\star\}_{i=1}^m$ once through equation \eqref{eq:QIF-offline-Q}, we propose a recursive updating procedure for online estimation and inference. In our proposed streaming updating procedure, let $\widetilde{\bbeta}_b$ denote the online estimator up to data batch $b$. We initialize $\widetilde{\bbeta}_1$ by the offline quadratic inference function estimator with the first data batch, i.e. $\widehat{\bbeta}_1=\arg \min_{\bbeta} \ Q_1(\bbeta)$. When data batches $\{\mD_{i,b}\}_{i=1}^m$ are collected, we update the previous estimator $\widetilde{\bbeta}_{b-1}$ to $\widetilde{\bbeta}_b$ using only summary statistics from previous data batches $\{\mD_{i,b-1}^\star \}_{i=1}^m$ and the raw data in the current data batch $\{\mD_{i,b}\}_{i=1}^m$. After completing the updating, individual-level data in $\{\mD_{i,b}\}_{i=1}^m$ is no longer accessible for the sake of storage. We can only access the updated estimate $\widetilde{\bbeta}_b$, and summary statistics are carried forward for future updating.

For clarity of exposition, we begin our derivation with two data batches $\mD_{i,1}$ and $\mD_{i,2}$ that are collected sequentially at time points $t_1$ and $t_2$ respectively, where $\mD_{i,2}$ arrives after $\mD_{i,1}$ for $i=1,\dots,m$. Following \cite{Qu2000}, the quadratic inference function estimator $\widehat{\bbeta}_1 = \arg \min_{\bbeta} \ Q_1(\bbeta)$ satisfies $\bS_1^\top ( \widehat{\bbeta}_1 ) \{\bV_1( \widehat{\bbeta}_1 ) \}^{-1} \bU_1 ( \widehat{\bbeta}_1 ) = \bzero$, where $Q_1(\bbeta)=\bU_1^\top( \bbeta ) \{\bV_1 ( \bbeta) \}^{-1} \bU_1 ( \bbeta )$. When $\{\mD_{i,2} \}_{ i=1 }^m$ arrives, we can obtain the offline quadratic inference function estimator $\widehat{\bbeta}_2^\star$ based on the cumulative dataset $\{ \mD_{i,2}^\star \}_{i=1}^m$ by solving the estimating equation $\bS_2^\star( \widehat{\bbeta}_2^\star )^\top \{\bV_2^\star ( \widehat{\bbeta}_2^\star )\}^{-1} \bU_2^\star ( \widehat{\bbeta}_2^\star ) = \bzero$, where each of the building blocks can be decomposed as follows:
\begin{align}
\bU_{i,2}^\star(\widehat{\bbeta}_2^\star) &= 
q^{t_2-t_1}
\bU_{i,1}(\widehat{\bbeta}_2^\star)
+ 
\begin{pmatrix}
\bU_{i,2}(\widehat{\bbeta}_2^\star)^{(1)} \\
\bU_{i,2}(\widehat{\bbeta}_2^\star)^{(2)} + \bU_{i,1,2}(\widehat{\bbeta}_2^\star) + q^{t_2-t_1} \bU_{i,2,1}(\widehat{\bbeta}_2^\star)
\end{pmatrix}, \nonumber \\
\bS_{i,2}^\star(\widehat{\bbeta}_2^\star) &= q^{t_2-t_1} \bS_{i,1}(\widehat{\bbeta}_2^\star) + 
\begin{pmatrix}
\bS_{i,2}(\widehat{\bbeta}_2^\star)^{(1)} \\
\bS_{i,2}(\widehat{\bbeta}_2^\star)^{(2)} + \bS_{i,1,2}(\widehat{\bbeta}_2^\star) + q^{t_2-t_1} \bS_{i,2,1}(\widehat{\bbeta}_2^\star)
\end{pmatrix}, \label{eq:statistics_decompose} \\
\bU_{2}^\star(\widehat{\bbeta}_2^\star) &= \sum_{i=1}^m \bU_{i,2}^\star(\widehat{\bbeta}_2^\star), \
\bV_2^\star(\widehat{\bbeta}_2^\star) = \sum_{i=1}^m \bU_{i,2}^\star(\widehat{\bbeta}_2^\star)\left[\bU_{i,2}^\star(\widehat{\bbeta}_2^\star)\right]^\top, \ 
\bS_2^\star(\widehat{\bbeta}_2^\star)=\sum_{i=1}^m \bS_{i,2}^\star(\widehat{\bbeta}_2^\star). \nonumber
\end{align}

Even though the quantities in equation \eqref{eq:statistics_decompose} admit recursive updating forms, plugging in the newly obtained estimator $\widehat{\bbeta}_2^\star$ into the old summary statistics, such as $\bU_{1}$, requires access to historical raw data $\{\mD_{i,1}\}_{i=1}^m$. This incurs both a data storage and a recomputation burden. To avoid reusing historical raw data, we do not carry out calculations retrospectively. To derive an online streaming estimation procedure, we take a first-order Taylor expansion of the terms $\bU_{1}( \widehat{\bbeta}_2^\star )$ and $\bS_1( \widehat{\bbeta}_2^\star )$ around $\widehat{\bbeta}_1$ and obtain
\begin{equation}\label{eq:statistics_online}
\begin{split}
\frac{n_1}{N_2}{\bU}_1(\widehat{\bbeta}_2^\star) &= \frac{n_1}{N_2}{\bU}_{1}(\widehat{\bbeta}_{1}) +\frac{n_1}{N_2}
{\bS}_{1}(\widehat{\bbeta}_1)(\widehat{\bbeta}_{1} - \widehat{\bbeta}_2^\star)+ O_p\left(\frac{n_1}{N_2}\|\widehat{\bbeta}_1 - \widehat{\bbeta}_2^\star\|^2 \right), \\
\frac{n_1}{N_2}{\bS}_1(\widehat{\bbeta}_2^\star) &= \frac{n_1}{N_2}{\bS}_{1}(\widehat{\bbeta}_1) + O_p\left(\frac{n_1}{N_2}\|\widehat{\bbeta}_1-\widehat{\bbeta}_2^\star \| \right). \\
\end{split}
\end{equation}
The error terms $O_p (n_1 \|\widehat{\bbeta}_1-\widehat{\bbeta}_2^\star\|/N_2 )$ and $O_p (n_1 \|\widehat{\bbeta}_1 - \widehat{\bbeta}_2^\star\|^2/N_2 )$ in equation \eqref{eq:statistics_online} may be asymptotically ignored if $N_2$ is large enough and $\|\bbeta_2-\bbeta_1\|=o(1)$. We drop these higher-order terms and propose an online streaming quadratic inference function estimator $\widetilde{\bbeta}_2$ as a solution to the estimating equation $\widetilde{\bS}_2^\top\widetilde{\bV}_2^{-1}\widetilde{\bU}_2=\bzero$, where $\widetilde{\bU}_2= q^{t_2-t_1}\bU_1(\widehat{\bbeta}_1)+ q^{t_2-t_1}\bS_1(\widehat{\bbeta}_1)(\widehat{\bbeta}_1- \widetilde{\bbeta}_2)+\bU_2(\widetilde{\bbeta}_2)$ is the adjusted score function and $\widetilde{\bS}_2 = q^{t_2-t_1}\bS_1 ( \widehat{\bbeta}_1 ) + \bS_2 ( \widetilde{\bbeta}_2 )$ and $\widetilde{\bV}_2 = \sum_{i=1}^m \widetilde{\bU}_{i,2} \widetilde{\bU}_{i,2}^\top$ are the aggregated negative gradient and sample variability matrix of $\widetilde{\bU}_2$, respectively. Therefore, we can update $\widehat{\bbeta}_1$ to $\widetilde{\bbeta}_2$ without reaccessing individual level data in $\{\mD_{i,1}\}_{i=1}^m$. Note that, in contrast to the independent case, the sample variability matrix does not take a linear aggregation form. In addition, we can solve for $\widetilde{\bbeta}_2$ via the Newton-Raphson algorithm. Specifically, at the $(r+1)$-th iteration, 
\begin{align*}
\widetilde{\bbeta}_2^{(r+1)} = \widetilde{\bbeta}_2^{(r)} + \left\{\left(\widetilde{\bS}_2^{(r)}\right)^\top \left(\widetilde{\bV}_2^{(r)}\right)^{-1} \widetilde{\bS}_2^{(r)}\right\}^{-1}
\left(\widetilde{\bS}_2^{(r)}\right)^\top \left(\widetilde{\bV}_2^{(r)}\right)^{-1} \widetilde{\bU}_2^{(r)},
\end{align*}
where 
\begin{align*}
\widetilde{\bU}_{i,2}^{(r)} &=
q^{t_2-t_1}\bU_{i,1}(\widehat{\bbeta}_1)+q^{t_2-t_1}\bS_{i,1}(\widehat{\bbeta}_1)(\widehat{\bbeta}_1-\widetilde{\bbeta}_2^{(r)}) \\
& +
\begin{pmatrix}
\bU_{i,2}(\widetilde{\bbeta}_2^{(r)})^{(1)} \\
\bU_{i,2}(\widetilde{\bbeta}_2^{(r)})^{(2)} + \bU_{i,1,2}(\widetilde{\bbeta}_2^{(r)}) + q^{t_2-t_1}\bU_{i,2,1}(\widetilde{\bbeta}_2^{(r)})
\end{pmatrix}. \\
\widetilde{\bS}_{i,2}^{(r)}&= q^{t_2-t_1}\bS_{i,1}(\widehat{\bbeta}_1)+
\begin{pmatrix}
\bS_{i,2}(\widetilde{\bbeta}_2^{(r)})^{(1)} \\
\bS_{i,2}(\widetilde{\bbeta}_2^{(r)})^{(2)} + \bS_{i,1,2}(\widetilde{\bbeta}_2^{(r)}) + q^{t_2-t_1}\bS_{i,2,1}(\widetilde{\bbeta}_2^{(r)})
\end{pmatrix}, \\
\widetilde{\bU}_{2}(\widehat{\bbeta}_2^{(r)}) &= \sum_{i=1}^m \widetilde{\bU}_{i,2}^{(r)}, \qquad
\widetilde{\bV}_2(\widehat{\bbeta}_2^{(r)}) = \sum_{i=1}^m \widetilde{\bU}_{i,2}^{(r)}\left[\widetilde{\bU}_{i,2}^{(r)}\right]^\top, \qquad
\widetilde{\bS}^{(r)}_2=\sum_{i=1}^m \widetilde{\bS}_{i,2}^{(r)}.
\end{align*}

Remarkably, due to the decomposition in \eqref{eq:offlineEE-pieces}, individual-level data in $\mD_1$ is not used except for the last observation $\{\bx_{i,n_1}, y_{i,n_1}\}_{i=1}^m$. This allows us to avoid storing historical raw data while still accounting for and leveraging dependence between data batches for more efficient inference.

Generalizing the above procedure to a general streaming data setting where we want to update $\widetilde{\bbeta}_{b-1}$ to $\widetilde{\bbeta}_b$, the online estimator $\widetilde{\bbeta}_b$ of $\bbeta$ is the solution to the incremental estimating equation
\begin{equation}\label{eq:incremental_EE}
\widetilde{\bS}_b^\top \widetilde{\bV}_b^{-1} \widetilde{\bU}_b = \bzero,
\end{equation}
where the building blocks are updated in a similar way to equation~\eqref{eq:statistics_online}: 
\[
\begin{aligned}
\widetilde{\bU}_{i,b}&= 
q^{t_b-t_{b-1}}\widetilde{\bU}_{i,b-1} + q^{t_b-t_{b-1}}\widetilde{\bS}_{i,b-1}(\widetilde{\bbeta}_{b-1}-\widetilde{\bbeta}_b)\\
& +
\begin{pmatrix}
{\bU}_{i,b}(\widetilde{\bbeta}_b)^{(1)} \\
{\bU}_{i,b}(\widetilde{\bbeta}_b)^{(2)} + \bU_{i,b-1,b}(\widetilde{\bbeta}_b) + q^{t_b-t_{b-1}}\bU_{i,b,b-1}(\widetilde{\bbeta}_b) \\
\end{pmatrix}, \\
\widetilde{\bS}_{i,b} & = 
q^{t_b-t_{b-1}}\widetilde{\bS}_{i,b-1} 
+
\begin{pmatrix}
{\bS}_{i,b}(\widetilde{\bbeta}_b)^{(1)} \\
{\bS}_{i,b}(\widetilde{\bbeta}_b)^{(2)} + \bS_{i,b-1,b}(\widetilde{\bbeta}_b) + q^{t_b-t_{b-1}}\bS_{i,b,b-1}(\widetilde{\bbeta}_b) \\
\end{pmatrix}, \\
\widetilde{\bU}_b & = \sum_{i=1}^m \widetilde{\bU}_{i,b}, \qquad
\widetilde{\bV}_b  = \sum_{i=1}^m \widetilde{\bU}_{i,b}\widetilde{\bU}_{i,b}^\top, \qquad 
\widetilde{\bS}_b =\sum_{i=1}^m \widetilde{\bS}_{i,b}. \\
\end{aligned}
\]

Solving equation \eqref{eq:incremental_EE} can be done via the Newton-Raphson algorithm with the $(r+1)$-th iteration taking the form:
\begin{align*}
\widetilde{\bbeta}^{(r+1)}_b&=\widetilde{\bbeta}^{(r)}_b + \left\{ \widetilde{\bS}^{(r)\top}_b \left( \widetilde{\bV}^{(r)}_b\right)^{-1} \widetilde{\bS}^{(r)}_b \right\}^{-1} \widetilde{\bS}^{(r)\top}_b \left( \widetilde{\bV}^{(r)}_b\right)^{-1} \widetilde{\bU}^{(r)}_b,
\end{align*}
where we don't need to access the entire raw dataset except for the observations in the current batch $\mD_{i,b}$ and the last observation in data batch $\mD_{i,b-1}$. Instead, we only use the previous estimate $\widetilde{\bbeta}_{b-1}$, as well as summary statistics $\{\widetilde{\bU}_{i,b-1},\widetilde{\bS}_{i,b-1}\}_{i=1}^m$ from historical data up to time point  $t_{b-1}$.

Finally, we propose an adaptive tuning procedure for selecting the weighting parameter $q$. Let $\mathcal{C}_q$ denote a candidate set for $q$. At updating time point $t_b$, we compute $\widetilde{\bbeta}_b(q)$ for all $q \in \mathcal{C}_q$, and choose $q$ that minimizes the contribution of the $b$th batch to the cumulative quadratic inference function up to batch $b$. Specifically, let
\begin{align*}
\bU_{i,b}(\widetilde{\bbeta}_b, q)&=
\begin{pmatrix}
{\bU}_{i,b}(\widetilde{\bbeta}_b)^{(1)} \\
{\bU}_{i,b}(\widetilde{\bbeta}_b)^{(2)} + \bU_{i,b-1,b}(\widetilde{\bbeta}_b) + q^{t_b-t_{b-1}}\bU_{i,b,b-1}(\widetilde{\bbeta}_b) \\
\end{pmatrix}\\
\bU_b(\widetilde{\bbeta}_b, q)&=\sum \limits_{i=1}^m \bU_{i,b}(\widetilde{\bbeta}_b, q), \quad \bV_b(\widetilde{\bbeta}_b, q)=\sum \limits_{i=1}^m \bU_{i,b}(\widetilde{\bbeta}_b, q) \bU_{i,b}(\widetilde{\bbeta}_b, q)^\top.
\end{align*}
We propose to select $q$ using
\begin{align}
q_b^{\textup{opt}} &= \arg \min \limits_{q \in \mathcal{\bC}_q} U_b(\widetilde{\bbeta}_b, q)^\top  \left\{\bV_b(\widetilde{\bbeta}_b, q)\right\}^{-1}
\bU_b(\widetilde{\bbeta}_b, q).
\label{e:q_opt}
\end{align}

\section{Large Sample Properties}
\label{s:asymptotics}

We establish large sample properties of our proposed online estimator $\widetilde{\bbeta}_b$ in equation \eqref{eq:incremental_EE} under the condition where $n_j$ is finite for every $j=1,\dots,b$, but the number of data batches $b\to\infty$. The technical difficulty arises from the fact that $n_j$ is finite and the convergence is driven by the number of iterative steps indexed by $b$. We first define population quantities of interest: let the sensitivity and variability matrices for batch $b$ be denoted by $\mathbb{S}(\bbeta(t_b))=n_b^{-1}\mathbb{E} [\bS_{i,b}(\bbeta(t_b))]=n_b^{-1}\mathbb{E}\left\{-\partial \bU_{i,b}(\by_i; \bX_i,\bbeta(t_b))/\partial\bbeta(t_b)^\top\right\}$ and $\mathbb{\bV}(\bbeta(t_b))=n_b^{-1}\mathbb{E}\left\{\bU_{i,b}(\by_i;\bX_i,\bbeta(t_b))\bU_{i,b}(\by_i; \bX_i,\bbeta(t_b))^\top\right\}$. 
We consider a set of assumptions given below.

\begin{enumerate}[label=(A\arabic*),leftmargin=*]
\item For participant $i=1,\dots,m$, the expectation $\mathbb{E}[n_b^{-1}\bU_{i,b}(\by_i;\bX_i,\bbeta)]=\bzero$ if and only if $\bbeta=\bbeta_b$. Furthermore, $\mathbb{E}[\|n_b^{-1}\bU_{i,b}(\by_i;\bX_i,\bbeta(t_b))\|^r]<\infty$, $r=1,2$ for all updating time points. \label{C2:unique}
\item The matrix ${\bH}_b=N_b^{-1}{\widetilde{\bS}}_b^\top{\widetilde{\bV}}_b^{-1}{\widetilde{\bS}}_{b-1}$ is positive-definite for $b\geq2$. \label{C5:pd_H}
\item For all considered updating time points, the function of the score vector $\bU_{i,b}^\star(\by_i;\bX_i, \bbeta(t_b) )$ is twice continuously differentiable in $\bbeta(t_b)$, and the sensitivity matrix $\mathbb{S}(\bbeta(t_b))=n_b^{-1}\mathbb{E}[ \bS_b(\mD_b;\bbeta(t_b))]$ is of full column rank.
\label{C3:differentiable}
\item \label{C6:phi_mixing} For every individual $i$, the vector $(y_{i,kj}-\mu_{i,kj})_{k,j=1}^{n_j,b}$ forms a $\rho$-mixing stochastic process. If $\rho(l)$ denote the mixing coefficients for $l=1,2,\ldots$, then $\sum_{l} \rho(l) <\infty$. 
\item \label{Assumption:varying-coefficient} The time-varying coefficient $\bbeta(t)$ is twice differentiable with respect to $t$ with bounded derivative. During the considered time period $\mathcal{T}$ defined on a compact set, $\sup_{t\in\mathcal{T}}\bbeta(t)$ is bounded. The adjacent batches share similar time-varying coefficients in the sense that 
$b\times \sup_{1\leq j\leq b}| \bbeta(t_j) - \bbeta(t_{j-1}) |<\infty$. 
\item \label{Assumption:asymptotic-regimes} The number of time points $N_b$ observed in the algorithm running period and the tuning parameter $q$ satisfy $-1/\log q\rightarrow 0$, and, as $b\rightarrow\infty$,  $N_b\rightarrow\infty$, $-N_b/\log q\rightarrow \infty$ and $\log N_b/\sqrt{-N_b/(\log q)^3}\rightarrow 0$.
\item For all considered updating time points, the variability matrix $\mathbb{V}\big(\bbeta(t_b)\big)$ is positive-definite.
\label{C4:positive-definite}
\end{enumerate}

Assumptions \ref{C2:unique}-\ref{C3:differentiable} and \ref{C4:positive-definite} are regularity assumptions required to establish asymptotic consistency and asymptotic normality of the generalized method of moments estimator \citep{Hansen1982}. The matrix $\bH_b$ in Assumption \ref{C5:pd_H} approximates the sample version of the covariance matrix of $\tbeta_b$ and its positive-definiteness is needed to ensure the feasibility of statistical inference. Assumption \ref{C6:phi_mixing} is a reasonable condition for temporal time series where dependence decays to zero with high polynomial rates over time. Assumption \ref{Assumption:varying-coefficient} is the smoothness condition commonly adopted in the literature on varying coefficient models \citep{fan1999statistical}. By restricting the maximal change of the coefficient $\bbeta(t)$ for adjacent time points during the batch updating period, we assume that the conditional distribution of the outcome given the covariates changes gradually over time. This is a reasonable expectation given our motivating wearable device application. Assumption \ref{Assumption:asymptotic-regimes} imposes restrictions on the smoothing parameter $q$ and the number of updates $b$. The smoothing parameter is chosen at a faster rate than in the standard nonparametric regression problems so that the smoothing bias vanishes, allowing our procedure to yield valid statistical inference. 

\begin{theorem}
\label{thm:consist}
Under Assumptions \ref{C2:unique}-\ref{Assumption:asymptotic-regimes}, the online estimator $\widetilde{\bbeta}_b$ given in \eqref{eq:incremental_EE} is consistent, namely $\widetilde{\bbeta}_b\overset{p}{\to}\bbeta_b$, as $N_b=\sum_{j=1}^b n_j\to\infty$.
\end{theorem}

\begin{theorem}\label{thm:normal}
Under Assumptions \ref{C2:unique}-\ref{C4:positive-definite}, the online estimator $\widetilde{\bbeta}_b$ given in \eqref{eq:incremental_EE} is asymptotically normally distributed, namely $\sqrt{-N_b/\log q}(\widetilde{\bbeta}_b-\bbeta_b)\overset{d}{\to}\mathcal{N}(\bzero,\mathbb{J}^{-1}(\bbeta_b))$, as $N_b=\sum_{j=1}^b n_j\to\infty$, where $\mathbb{J}(\bbeta_b)=\mathbb{S}^\top(\bbeta_b)\mathbb{V}^{-1}(\bbeta_b)\mathbb{S}(\bbeta_b)$ is the Godambe information matrix.
\end{theorem}

The proofs of these theorems are provided in the Supplementary Material. Importantly, the asymptotic covariance matrix of the online estimator $\tbeta_b$ given in Theorem~\ref{thm:normal} is exactly the same as that of the offline estimator $\widehat{\bbeta}_b^\star$. This implies that the proposed online estimator achieves the same asymptotic distribution as its offline counterpart. Without reaccessing previous historical data, we use aggregated inferential matrices $\widetilde{\bS}_b$ and $\widetilde{\bV}_b$ to estimate the Godambe information matrix by $\widetilde{\mathbb{J}}(\bbeta_b)=\{-\log q\widetilde{\bS}_b^\top\widetilde{\bV}_b^{-1}\widetilde{\bS}_b / N_b\}$. Then the estimated variance of $\widetilde{\bbeta}_b$ is  $\widetilde{\text{var}}(\tbeta_b)=\{-\log q\widetilde{\mathbb{J}}(\bbeta_b) /N_b\}^{-1} =\{ \widetilde{\bS}_b^\top\widetilde{\bV}_b^{-1}\widetilde{\bS}_b\}^{-1}.$

\section{Simulations}
\label{s:simulations}

\subsection{Simulation setting}

In this section, we examine the finite sample performance of the proposed streaming estimator $\widetilde{\bbeta}_b$ in \eqref{eq:incremental_EE} and its estimated covariance $\widetilde{\text{var}}(\tbeta_b)$ in Theorem \ref{thm:normal} through simulations. In all numerical experiments, we consider a sequence of equally spaced updating time points $t_1,t_2\dots,t_b$ at which the data batches are collected. We assume at a certain time $t_j$, a batch of $n_j$ observations are collected, and the weight applied to observations in this batch is $q^{b-j}$. All simulations are run on a standard Linux cluster with one central processing unit and 1 gigabyte of random-access memory. We consider two sets of simulations in the linear and logistic models in the main manuscript. An additional simulation in the Poisson regression setting included in the Supplementary Material corroborates the findings from the linear and logistic regression simulations. This Poisson simulation mimics the data analyzed in Section \ref{s:data}. 

We simulate $b=200$ batches of sizes $n_j=20$. Covariates $\bX_{i,j} = ( \bx^\top_{i, kj} )_{k=1}^{n_j}$ consist of an intercept and two longitudinal covariates independently simulated from a standard $n_j$-variate Normal distribution. We allow the effect of the first covariate to be batch-heterogeneous. The tuning parameter $q^{opt}_j$ is selected using equation \eqref{e:q_opt} from the candidate set $\mathcal{C}_q=\exp(-ab^{0.3})$ where $a$ is a sequence of $20$ evenly spaced scalars in $[0.1,1]$. We report the root mean squared error (\RMSE), empirical standard error (\ESE), bias (\BIAS), 95\% confidence interval coverage (\CP) and length (\LEN) of $\widetilde{\bbeta}_b$ in the last batch averaged over 500 simulation replicates. In the Supplementary Material, we include plots of the estimated heterogeneous regression coefficient in all batches and simulations to visually confirm that estimation consistency is stable across batches.

For comparison to our streaming estimator $\widetilde{\bbeta}_b$, we consider an online estimator that consists of our streaming estimator derived using a working independence assumption with the quadratic inference function decomposition in Sections \ref{ss:offline} and \ref{ss:online}, i.e. only one basis matrix $\bM_1$ corresponding to the identity matrix; it is computed using the same selection procedure for $q^{opt}_j$ as our streaming estimator.

\subsection{Linear regression simulation results}

In the first simulation, we consider the linear regression setting $E(y_{i,kj}\mid x_{i,kj})=\bx^\top_{i,kj} \bbeta_j$, where $\bx_{i,kj}$ is the $p$-dimensional covariate vector for participant $i$ at the $k$th observation in batch $j$, $k=1, \ldots, n_j$, $j=1, \ldots, b$, $i=1, \ldots, m$ with $m=100$. We set covariate effects $\bbeta_j=\{0.2, \sin(2\pi j/b), 0.5\}^\top$. We simulate $\by_i = ( y_{i,kj} )_{k,j=1}^{ n_j,b }$ from a normal distribution with mean $\bX_{i,j} \bbeta_j$ and covariance $\bSigma$ jointly over $b$ batches to control the correlation structure across batches, where $\bSigma$ corresponds to a first-order autoregressive covariance structure with variance $\sigma^2=4$ and correlation $\rho=0.8$. We report the \RMSE, \ESE, \BIAS, \CP~ and \LEN~of $\widetilde{\bbeta}_b$ in Table \ref{t:linear:sim}.

From Table \ref{t:linear:sim}, the bias \BIAS~of $\widetilde{\bbeta}_b$ appears negligible and the \RMSE~of $\widetilde{\bbeta}_b$ approximates the \ESE, supporting the consistency and asymptotic normality results of Theorem \ref{thm:consist} and \ref{thm:normal} when $b$ is finite. We observe appropriate 95\% confidence interval coverage, supporting the inferential properties of our estimator in finite $b$ settings.

Simulation metrics for the independent online estimator are presented in the Supplementary Material. Despite achieving nominal coverage, the independent online estimator has 95\% confidence intervals for covariates $\bX_1$ and $\bX_2$ that are, on average, approximately $20\%$ longer. Indeed, by accounting for dependence, the streaming estimator is more efficient, highlighting the statistical efficiency gains of our approach.

\begin{table}[h]
\centering
\caption{Simulation metrics for the streaming estimator of the first simulation: $m=100$, $b=200$ batches of sizes $n_j=20$. \label{t:linear:sim}}
\begin{tabular}{lrrrrr}
Covariate & \RMSE$\times 10^{-1}$ & \ESE$\times 10^{-1}$ & \BIAS$\times 10^{-3}$ & \CP & \LEN$\times 10^{-1}$ \\
Intercept & 1.24 & 1.24 & 4.30 & 0.95 & 6.72 \\ 
$\bX_1$ & 0.25 & 0.25 & 0.44 & 0.97 & 1.67 \\ 
$\bX_2$ & 0.25 & 0.25 & -0.77 & 0.94 & 1.39 \\
\end{tabular}
\end{table}

\subsection{Logistic regression simulation results}

In the second simulation, we consider the marginal logistic regression setting $E(y_{i,kj}\mid \bx_{i,kj})=\exp(\bx^\top_{i,kj} \bbeta_j)/\{1+\exp(\bx^\top_{i,kj}\bbeta_j) \}$, where $\bx_{i,kj}$ is the $p$-dimensional covariate vector for participant $i$ at observation $k$ in batch $j$, $k=1, \ldots, n_j$, $j=1, \ldots, b$, $i=1, \ldots, m$ with $m=100$. We set covariate effects $\bbeta_j=\{0.2, 4j(1-j/b)/b, 0.5\}^\top$. We simulate $\by_i = ( y_{i,kj} )_{ k,j=1}^{ n_j,b }$ from a multivariate Bernoulli distribution with mean $\exp(\bx_{i,kj}^\top\beta_j) / \{1 + \exp( \bx_{i,kj}^\top \beta_j ) \}$ using the \verb|SimCorMultRes| R package with latent first-order autoregressive covariance structure with variance $\sigma^2=4$ and correlation $\rho=0.8$. We report the \RMSE, \ESE, \BIAS, \CP~and \LEN~of $\widetilde{\bbeta}_b$ in Table \ref{t:logistic:sim}. The \RMSE~approximates the \ESE~well and the \BIAS~is negligible. We observe appropriate 95\% confidence interval coverage. We again conclude that the asymptotic properties of $\widetilde{\bbeta}_b$ in Theorems \ref{thm:consist} and \ref{thm:normal} appear to hold in finite $b$ settings. This simulation supports the use of our streaming estimator in marginal generalized linear models.

Simulation metrics for the independent online estimator are presented in the Supplementary Material. The independent online estimator has 95\% confidence intervals for covariates $\bX_1$ and $\bX_2$ that are, on average, approximately $5\%$ longer, corroborating our findings from the first simulation that our streaming estimator results in more efficient inference by leveraging the dependence structure of the outcomes. We highlight here that it is difficult to simulate Bernoulli outcomes with exact autoregressive correlation structure, so that the apparently small gain in efficiency is likely due to this imprecision and would be more significant for outcomes with an exact autoregressive correlation structure.

\begin{table}[h]
\centering
\caption{Simulation metrics for the streaming estimator in the second simulation $m=100$, $b=200$ batches of size $n_j=20$. \label{t:logistic:sim}}
\begin{tabular}{lrrrrr}
Covariate & \RMSE$\times 10^{-1}$ & \ESE$\times 10^{-1}$ & \BIAS$\times 10^{-3}$ & \CP & \LEN$\times 10^{-1}$ \\ 
Intercept & 1.04 & 1.04 & 1.42 & 0.95 & 4.97 \\ 
$\bX_1$ & 0.36 & 0.36 & 0.93 & 0.97 & 1.71 \\ 
$\bX_2$ & 0.46 & 0.46 & 3.57 & 0.96 & 2.09 \\
\end{tabular}
\end{table}

\section{Analysis of Accelerometer Data}
\label{s:data}

\begin{figure}[H]
\centering
\includegraphics[width=\textwidth]{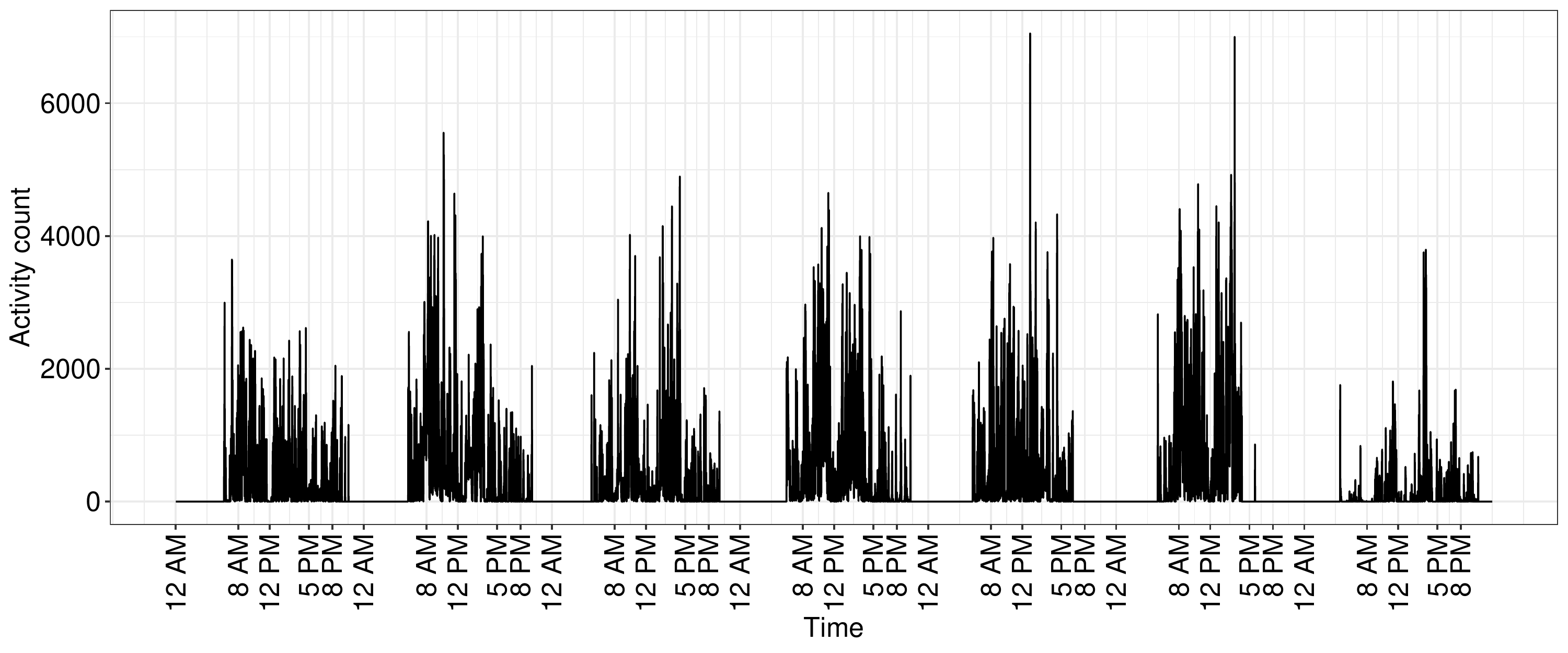}
\caption{Outcome data for one study participant.\label{f:NHANES-data}}
\end{figure}

We return to the motivating example introduced in Section \ref{s:introduction}. \cite{Leroux-etal} and the accompanying R package \verb|rnhanesdata| provide the data and a detailed pipeline for processing and analyzing the National Health and Nutrition Examination Survey accelerometry data. Following their proposed preprocessing pipeline, we analyze accelerometer activity counts for $m=1642$ study participants. For each participant $i=1, \ldots, m$, the outcome consists of activity counts for 1440 minutes per day for 7 days, for a total of $10080$ longitudinal outcomes. Outcome data for one participant are visualized in Figure \ref{f:NHANES-data}. 

Using our proposed streaming estimator, we investigate the association of activity counts with diseases adjusting for covariates through the Poisson regression model $\log \{ E( y_{i,kj} \mid \bx_{i,kj}) \}= \alpha + \bx_{i,kj}^\top \bbeta_j$, $k=1, \ldots, n_j$, $j=1, \ldots, b$, with batches of size $n_j=120$ (corresponding to two hours), $j=1, \ldots, 84$, with $q^{opt}_j=10^{-5}$ considered fixed. Covariates $\bx_{i,kj}$ consist of body mass index (\BMI, mean=$28.7$, standard deviation=$5.7$), coronary heart disease (\CHD, $0$=no, $1$=yes), coronary heart failure (\CHF, $0$=no, $1$=yes), cancer ($0$=no, $1$=yes), stroke ($0$=no, $1$=yes), diabetes ($0$=no, $1$=yes), sex ($0$=male, $1$=female), education ($0$=high school or less, $1$=more than high school), and self-reported mobility difficulties ($0$=none, $1$=some).

Estimated regression coefficients with $95$\% confidence intervals are visualized in Figure \ref{f:NHANES-trace}; we present trace plot of $p$-values and estimated effect sign across time in the Supplementary Material. The trends observed over time across the week are consistent with our intuition: we would expect to see cyclical associations that reflect diurnal and nocturnal activity patterns. This is especially evident for the intercept, which appears to capture intrinsic variations of activity over time that are not captured by our covariates. On the whole, covariate effects are primarily negative across the week with the exception of education. Indeed, it is not surprising that diseases such as coronary heart disease and coronary heart failure are negatively associated with physical activity. Women also appear to be less physically active than men throughout the week. Moreover, the magnitude of the sex effect is larger in the early morning, indicating that males are physically more active than females during these time periods. 

Trace plots of $p$-values are also useful for identifying time periods where a certain disease or confounder may be more important. Unsurprisingly, participants with mobility difficulties are significantly less physically active than those without throughout most of the day, except around 12AM when participants are presumably asleep. There appears to be a suggestive negative association between coronary heart disease and activity during the night, suggesting that participants with coronary heart disease may be more restless during sleep. Sex appears to be more strongly associated with physical activity in the mornings than in the afternoons.

The Poisson simulation setting in the Supplementary Material supports our discussion of significance levels in this data analysis. Nonetheless, we recommend caution when interpreting confidence intervals that have been computed multiple times for different batches because Type-I errors may not be well controlled due to multiple testing. In practice, $\alpha$ spending functions \citep{DeMets-Lan} may be of use for inference at multiple time points. 

\begin{figure}[h!]
\centering
\includegraphics[width=\textwidth]{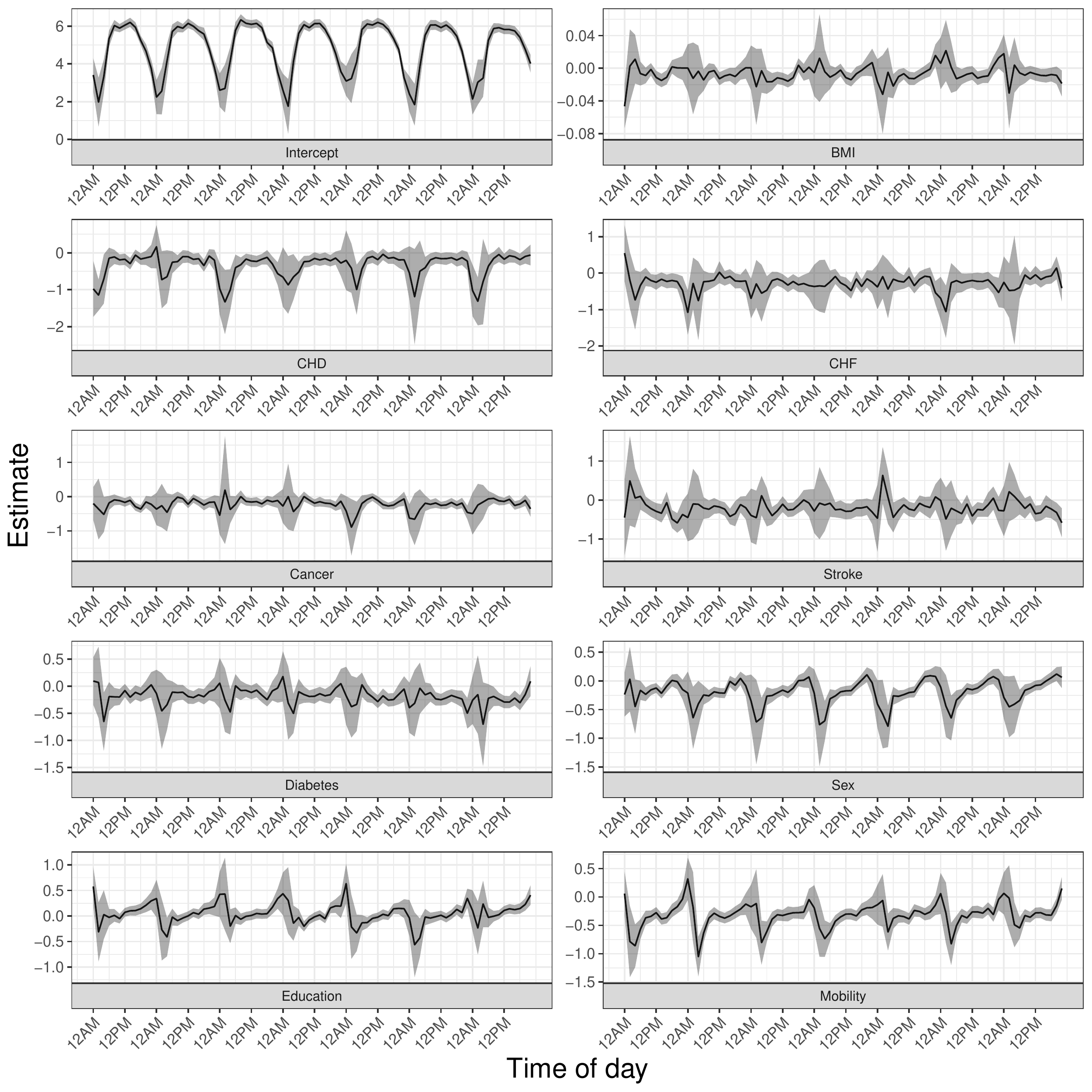}
\caption{Trace plot of estimated covariate effects across seven days, with $95$\% confidence intervals. \label{f:NHANES-trace}}
\end{figure}

\section{Discussion}

Despite the ubiquity of the first-order autoregressive process in longitudinal data analysis, other dependence structures may be of interest. The formulation of the extended score function in \eqref{e:QIF-offline-U} depends in general on the cumulative batches and does not admit a decomposition as in \eqref{eq:offlineEE-pieces} depending only on the previous data batch. This problem merits future consideration.

\appendix

\bibliographystyle{apalike}
\bibliography{bibliography-20210107}

\end{document}